\documentclass[twocolumn,preprintnumbers,amsmath,amssymb]{revtex4}
\usepackage{amsmath,epsfig,subfigure}
\usepackage{graphicx}
\usepackage{dcolumn}
\usepackage{bm}

\newcount\EquaNo \EquaNo=0
\def\newEq#1{\advance\EquaNo by 1 #1=\EquaNo}
\newcount\TablNo \TablNo=0
\def\newTabl#1{\advance\TablNo by 1 #1=\TablNo}
\newcount\FigNo \FigNo=0
\def\newFig#1{\advance\FigNo by 1 #1=\FigNo}
\newcount\ChapNo \ChapNo=0
\def\newCh#1{\advance\ChapNo by 1 #1=\ChapNo}
\begin {document}

\title { Methods for parallel simulations of surface reactions}

\author{S.V. Nedea}
\altaffiliation {Department of Mathematics and Computer Science, Technical University of Eindhoven}
\email{silvia@win.tue.nl}
\author{J.J. Lukkien}
\altaffiliation{
Department of Mathematics and Computer Science, Technical University of Eindhoven}
\author{A.P.J. Jansen}
\altaffiliation{
 Department of Chemical Engineering, Technical University of Eindhoven}
\author{P.A.J. Hilbers}
\altaffiliation{
Department of Mathematics and Computer Science, Technical University of Eindhoven}


\begin {abstract}
 We study the opportunities for parallelism for the simulation of surface
reactions. We introduce the concept of a {\it partition} and we give new simulation
methods based on Cellular Automaton using partitions.
We elaborate on the advantages and disadvantages of the derived
algorithms comparing with Dynamic Monte Carlo algorithms.
  We find that we can get very fast simulations using this approach trading
accuracy for performance and we give experimental data for the simulation
of Ziff model. 
\end{abstract}

\maketitle

\section*{1. Introduction }

   Understanding chemical reactions that take place on catalytic surfaces is
of vital importance to the chemical industry. The knowledge can be used to 
improve catalysts thus yielding better reactors. In addition, improving the
catalytic process in engines can lead to improvement of environmental conditions.
Dynamic Monte Carlo simulations have turned out to be a powerful tool to model
and analyse surface reactions~\cite{b_g_s_j, jansen, lukkien,
gelten_santen_jansen, binder, silvia, gelten_all}. 
In such a DMC model, the surface is described as a two-dimensional lattice of 
sites that can be occupied by particles. The simulation consist of changing 
the surface over time by execution of reactions, which are merely modifcations of 
small neighborhoods.
Simple models can be simulated easily; however, for more complicated models and
for larger systems computation time and memory use become very large. One way of
dealing with this is to use parallelism.  

   There are roughly three ways to introduce parallelism in simulations.
   First one may choose an existing simulation algorithm and exploit inherent
concurrency. This boils down to a parallelization effort which is by no 
means straightforward but limited by the properties of the algorithm one starts
with. General parallel methods such as the optimistic method Time Warp or the
pessimistic methods suggested by Misra and Chandy can be
used~\cite{chandy_misra, fujimoto, reed_malony}. If
the model (and, hence, the algorithm) does not admit much concurrency one may 
try to change the model. The penalty in many cases will be that such a model is
less accurate but this may be acceptable. Third, the necessary statistics may
be obtained from the averaging of a large number of small, independent simulations 
(this is, in fact, an instance of the second approach).
The mathematical model for DMC simulations of surface reactions is the Master
Equation (~\cite{kampen}, see also section 2). The algorithms based on it are rather 
sequential in nature and do not admit an optimistic parallel simulation method 
like Time Warp. This is because each reaction disables many others and, hence,
an optimistic method would result in frequent roll-back. Successful parallel
methods therefore change the model.

The most frequently used alternative model is that of a Cellular Automaton (CA) 
model. The
CA model~\cite{levermore_boghosian} is inherently parallel: all sites of the lattice
are updated 
simultaneously according to the state of the neighboring
sites. The time is 
incremented after updating synchronously all
the sites of the lattice, such that
 the state of a cell at time $t$ depends on the state of the neighboring sites at
 time
$t-1$. Each site is updated according to a transition rule
dependent on the
 state of the neighboring sites.

 There are two problems with the CA. First, all patterns (i.e., neighborhood
occupations) are treated on the same footing. In practice, however, these patterns 
represent different reactions that proceed with a different speed. 
The solution is to use a so-called non-deterministic CA (NDCA) in which execution 
of reaction is performed with a probability dependent on its rate.
The second problem concerns the occurence of conflicts.
 These conflicts must be resolved but
they make it impossible to have a synchronous update.

 In this paper we introduce and study the partitioned CA model~\cite{weimar,
worsch}. Each site of the lattice belongs to a set of a partition. These sets are chosen such that
reactions concerning sites within the same set have no conflicts. If we have, 
for  example, a von Neumann neighborhood~\cite{toffoli_margolus}, the sites can be distributed
over five sets with this property. The result is that updates in the same 
partition can be done simultaneously.

 The paper is organized as follows. In section 2 we give the mathematical
description of the simulation model. DMC simulation methods and algorithms
are introduced in section 3. In section 4 we give the CA and its simulation 
algorithm. Section 5 introduces the idea of partitioning worked out into
several alternatives. Section 6 contains some experimental data. We end with
some conclusions.

\section*{2. The Model}

\begin{figure}[h]
\centering
\subfigure {\epsfig {figure=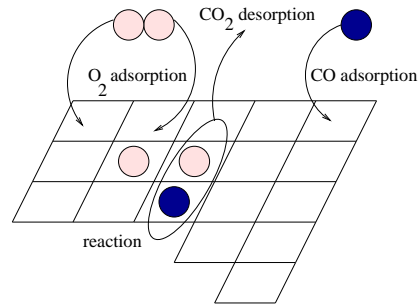, width=5.5cm} }
\caption { Reactions on a surface modelled by a two-dimensional lattice,
with two types of particles, ${\rm O}$(light colored) and ${\rm CO}$(dark colored).}
\label{Ziff_model}
\end {figure}

The model comprises the surface, the particles (molecules and atoms) and the reactions that take place. 
A picture of a model system is found in Fig.~\ref{Ziff_model}.
\\
\begin{center}{\it Lattice and set of states - $\Omega$, $D$}
\end{center}
 We model the surface by a two-dimmensional lattice $\Omega$
of $N$=$L_0$ x $L_1$ sites. 
 Every site takes a value from a set $D$, the domain of particle types. $D$ is a
finite set and generally contains an element to indicate that a site
is vacant, $D$=$\{*, {\rm A}, {\rm B}, \ldots \}$, where * stands for an
empty site.
  A complete assignment of particles to sites is called a
{\it system state} or {\it configuration}. Hence a configuration is a function from
$\Omega$ to $D$.

\begin{center}{\it The set of reaction types - $\cal T$}
\end{center}

A system state can be changed into a next state through a reaction. Such a
reaction is an instance of a {\it reaction type} at a specific site $s$. Occurrence
of a reaction means that the particles in a small neighborhood of $s$ are
replaced by new particles. A reaction can only occur when it is enabled,
 i.e., when the neighborhood is occupied according to some specific pattern. 
A reaction type therefore specifies
\begin{enumerate}
\item the neighborhood that should be considered for a given site,
\item the pattern needed for the reaction type in the neigborhood (the {\it source pattern}),
\item the pattern that will result from occurrence of the reaction (the {\it target pattern}).
\end{enumerate}

 A reaction type therefore is a function that, when applied to a specific site
$s$ yields a collection of triples of the form ({\it site, source, target}).
 The collection specifies the neighborhood and the source and target patterns. 
More precisely, a reaction type $Rt$ is a function from $\Omega$ to the
subsets of $\Omega$ x $D$ x $D$, ${\cal P}(\Omega$ x $D$ x $D)$.

 For $t$ $\in$ $Rt(s)$ we call the first component $t.site$, the second 
component $t.src$ and the third component $t.tg$. We will loosely refer to the set
 of first components as the neighborhood of $Rt$, the set of second components 
as the source pattern and the set of third components as the target pattern.
 The neighborhood of $Rt$ is a function, $Nb_{Rt}$: $\Omega$ $\rightarrow$ ${\cal P}(\Omega)$,
that, when applied to a site $s$, yields a collection of sites from
$\Omega$. This function must have the following two properties
\begin{enumerate}
\item  the neighborhood of a site $s$ includes $s$: $s \in Nb(s)$.
\item  translation invariance: the neighborhood looks the same for all sites
such that for any site $t$, $Nb(s+t)=Nb(s)+t=Nb(t)+s$.
\end{enumerate}
We can now also define when a  reaction type is enabled. A reaction type $Rt$ is
 enabled at site $s$ in state $S$ when for all $t$ in $Rt(s)$, $S(t.site) =
t.src$, i.e., when the source pattern matches at $s$ in $S$.

   Usually, there are many  reaction types and in a certain state many of them are
 enabled. A simulation proceeds by repeatedly selecting an enabled reaction 
 and executing it. The new state that results from executing an enabled
 reaction type $Rt$ in state $S$ at a site $s$ is defined as follows. Call this new
 state $S^{\prime}$ then 
\begin{enumerate}
\item $S^{\prime} (z) = S(z)$ if $z\notin Nb_{Rt}(s)$, i.e. $z$ does not occur in the neighborhood of
$Rt(s)$, 
\item for any $t$ in $Rt(s)$, $S^{\prime}(t.site) = t.tg$.
\end{enumerate}
\begin{center} {\it Reaction Rates}
\end{center}

  Each reaction type $Rt$ has a rate constant $k$ associated with it, which
is the probability per unit time that the reaction of type $Rt$ occurs.
 This rate constant depends on the temperature, expressed through an Arrhenius
expression $k={\nu}\exp{\left({-E}\over{k_BT}\right)}$ where
$E$ is the activation energy and $\nu$ the pre-exponential factor.

\begin{center} {\it Example}
\end{center}

 We give a simple example where we use the defined model to describe 
the surface reactions in a simple system.
 In this system~\cite{ziff}, particles of type $\rm CO$ (carbon monoxide) may adsorb on
the sites of a square lattice. Particles of type $\rm O$ (oxygen) may do
so as well, but in the gas phase $\rm O$ is diatomic.
 The adsorption of $\rm CO$ has rate constant $k_{\rm CO}$ and the dissociative
adsorption of $\rm O_2$ has rate constant $k_{\rm O_2}$. $\rm O_2$ adsorbs at
adjacent vacant sites.
 Adsorbed $\rm O$ and $\rm CO$ particles form $\rm CO_2$ and desorb (see
Fig. ~\ref{Ziff_model}).
The rate of formation and desorption of $\rm CO_2$ is $k_{\rm CO_2}$.
 The surface is modelled by a lattice of sites $\Omega$=$L_0$ x $L_1$, and 
$D$=$\{$*, $\rm CO$, $\rm O$$\}$.
 The reactions are modelled by seven  reaction types that we give 
in Table~\ref{tab:Table1}.
\begin {table*}
\begin {center}
\begin{tabular}{|l|p{2.2cm}p{2.2cm}p{2.3cm}p{2.5cm}|}
\hline   & 0 & 1 & 2 & 3\\
\hline   $Rt_{\rm CO+O}$& \{($s$,CO,*), ($s$+(1,0),O,*)\} & \{($s$,CO,*), ($s$+(0,1),O,*)\} & \{($s$,CO,*), ($s$+(-1,0),O,*)\} & \{($s$,CO,*), ($s$+(0,-1),CO,*)\}\\
\hline   $Rt_{\rm O_2}$ & \{($s$,*,O), ($s$+(1,0),*,O)\}   & \{($s$,*,O), ($s$+(0,1),*,O)\} & & \\
\hline   $Rt_{\rm CO}$  & \{($s$,*,CO)\}    &     & &  \\
\hline
\end{tabular}
\end{center}
\caption {\label{tab:Table1} All the transformations of the reaction types
that model the reactions of the system in Fig.~\ref{Ziff_model} applied on a site $s$. Notice that
$Rt_{\rm CO+O}$ has four versions corresponding to the four possible orientations of the pattern.
$Rt_{\rm O_2}$ has only two.}
\end{table*}

 For $\rm CO$ adsorption there is one reaction type, for $\rm O_2$ adsorption
there are two reaction types because two orientations are possible when the
$\rm O_2$ molecule dissociates. Finally, the $\rm CO_2$ formation and
desorption is modelled by four reaction types.

\section*{3. Simulations}

 Using different simulation techniques we want to study the evolution of the
defined model over time. This means that we repeatedly execute enabled
reactions on the lattice according to a simulation algorithm.
 There are basically two different approaches to simulate lattice
models: Dynamic Monte Carlo (DMC) and Cellular Automata (CA).

\subsection*{Dynamic Monte Carlo}

 Dynamic Monte Carlo methods are used to simulate the behavior of
stochastic models over time. The stochastic models of physical systems are based
on a Master Equation (ME) of the form 

\begin{equation}
  {dP(S,t)\over dt}
  =\sum_{S^{\prime}}
   \left[k_{SS^{\prime}}P(S^{\prime},t)-k_{S^{\prime}S}P(S,t)\right].
\end{equation}

 In this equation $P(S,t)$ denotes the probability to find the system in a
configuration $S$ at time t, $k_{SS^{\prime}}$ is the reaction probability
of the reaction that transfers $S^\prime$ into $S$. Such a reaction is an
instance of a reaction type at a particular location.

A simulation according to 
this equation is essentially a discrete event simulation
with continuous time, where the events are the occurences of reactions.
When we assume that the rate constants do not change over time, occurrence of each event has
a negative exponential distribution~\cite{jansen, lukkien}.
 In an abstract way, a simulation then consists of a repetition of the following
steps
\begin{enumerate}
\item Determine the list of enabled reactions.
\item Select the reaction that is the next and the time of its occurence.
\item Adjust the time and the lattice.
\item Continue from 1.
\end{enumerate}

 There are many different ways to organize these steps resulting in
different algorithms. The taxonomy by Segers~\cite{segers} mentions not less
than 48 DMC algorithms.
 One of these algorithms, called the Random Selection Method (RSM) has
become quite popular in the literature because of its simple
implementation. It reads as follows
\begin{tabbing}
\hspace{0.2cm} \= set time to 0;\\
               \> repeat\\
               \> ~~ \= 1. select a site $s$ randomly with probability $1/N$;\\
               \>    \> 2. select a reaction type $i$ with probability $k_i$/K;\\
               \>    \> 3. check if the reaction type is enabled at $s$;\\
               \>    \> 4. if it is, execute it;\\
               \>    \> 5. advance the time by drawing from\\
               \>    \> ~~ $\left[1-\exp(-NKt)\right]$;\\
               \> until simulation time has elapsed;\\
\end{tabbing}
 We denoted with $K$ the sum of the rate constants of the reaction types,
i.e., $K=\sum_i{k_i}$.
A single iteration of this algorithm is called a {\em trial} as no success is
guaranteed.
In the literature the connection with real
time (step 5) is usually not mentioned. 
Since the method decouples the notion of time
completely from the simulation (the time increment is drawn from a fixed
distribution) there is a tendency to measure time in Monte Carlo
steps. One MC step corresponds to one trial of a reaction type per lattice
site on average, i.e., one MC step is $N$ trials.

 This definition of MC steps also allows comparison between time-based simulation
techniques and MC step-based techniques. Thus, one MC step is equivalent to
$ \sum_{j=0}^{N-1}{T_j},$
where the time increment $T_j$ is selected from the fixed distribution 
$1-\exp{(-NKt)}$.
 This value can be drawn at the end of an MC step or at the end of the entire
simulation. Without using the negative exponential distribution, method RSM
can also be regarded as a time discretization of the ME. The time step is
then $1/NK$.

  In this paper we apply method RSM because it is a DMC method that is very similar
to a cellular automaton.
 RSM is a purely sequential algorithm that is not well suited for
parallelism. In ~\cite{segers, segers1}, Segers et al. investigated how parallelism may be
used in the simulation of surface reactions. The goal was to see whether
simulations that are too large to run on a sequential computer may run on a
parallel computer. He proposed an approach in which coherent parts of the lattice, so-called chunks,
are assigned to a number of processors. The chunks are then simulated using
RSM. When reactions are across
multiple lattice chunks, state information has to be exchanged. Thus,
synchronization and communication techniques are required for the chunk-boundaries.
 His investigation shows that for parallel simulations, the overhead of
the parallel algorithm is considerable because of the high communication latency
of parallel computers. 
In order to get a significant increase in speed over the sequential algorithm the 
amount of work on each processor must be large compared to the amount of communication.
This trade-off is given by the volume/boundary ratio of the blocks.
In this paper we explore another direction, viz., 
a parallel simulation algorithm that
{\em approximates} the kinetics of the ME, thus trading accuracy for performance.

\section*{4. Cellular Automaton}

  Cellular Automata are a powerful tool to simulate physical and chemical
systems of a high level of complexity.
 The CA approach is discrete in space and time. In a CA, all sites can make a
reaction in each step of the simulation.
For a standard Cellular Automaton the notion of real time
is discarded: the decision of whether to make a reaction is
based only on information local to that site. As a result, a
slowly evolving reaction at some part of the lattice, has the same
probability to occur as a fast reaction at another place. In order
to introduce the real-time dynamics, the decision of whether to
make a reaction is taken with a probability that depends on the
rate constant resulting in a Non-Deterministic Cellular Automaton (NDCA).
This NDCA resembles the RSM method the best. A NDCA algorithm consists of the following steps

\begin{tabbing}
\hspace{0.2cm} \= for each step\\
 \> ~~ \= for each site $s$\\
 \>   \> ~~\= 1. \= select a reaction type $i$ with probability $k_i/K$;\\
 \>   \>   \> 2. \= check whether the reaction is enabled at $s$;\\
 \>   \>   \> 3. if it is, execute it;\\
 \>   \>   \> 4. advance the time;\\
\end{tabbing}


 In RSM and NDCA the mechanism of selecting a site is different
and generates deviations in the simulation results. This happens because in a NDCA, each site is always selected during a step,
while in RSM there is a non-zero probability that a site is chosen twice or more in a 
succesion during one simulation step.
 This difference in selecting a site introduces biases in the rates of
the reactions and causes NDCA to give degenerate results for some
systems (Ising models, Single-File models, etc.)~\cite{vichniac}.

\begin{figure}
\centering
\subfigure {\epsfig {figure=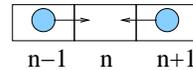, width=2.5cm} }
\caption { Conflict for simulating reactions that affect neighboring sites
(such as diffusion) in a CA model. Both particle can jump to
the empty site $n$ during the same step.}
\label{diff_conflict}
\end {figure}

  The CA approach is inherently parallel and all lattice sites can be updated
simultaneously in one single time step of the simulation. However,
parallelism also introduces conflicts for reactions that may disable each other.
Consider, for instance, a diffusion model involving two sites. A particle at 
site $n$ can jump to one of the neighboring sites if this is an empty site, 
while another particle from the same neighborhood could jump as well to the 
same empty site (see Fig.~\ref{diff_conflict}).
Failure to deal with this may lead to the violation of the physical laws.
 We avoid erroneous simulations by adjusting the simulation model in
order to avoid the conflicts that arise as a consequence of the parallel
execution.

\section*{5. Cellular Automaton with partitions}

  In the literature, the problem of the conflicts in parallel simulations is
solved using Block Cellular Automata (BCA).
 The BCA use the concept of partitioning. The sites of a CA are partitioned
into a regular pattern of blocks that cover the whole space and do not
overlap. A step is then applied at the same time and independently
to each block.
 In the next step, the block boundaries are shifted such that
the edges occur at a different place.

\begin{figure}[h]
\centering
\subfigure {\epsfig {figure=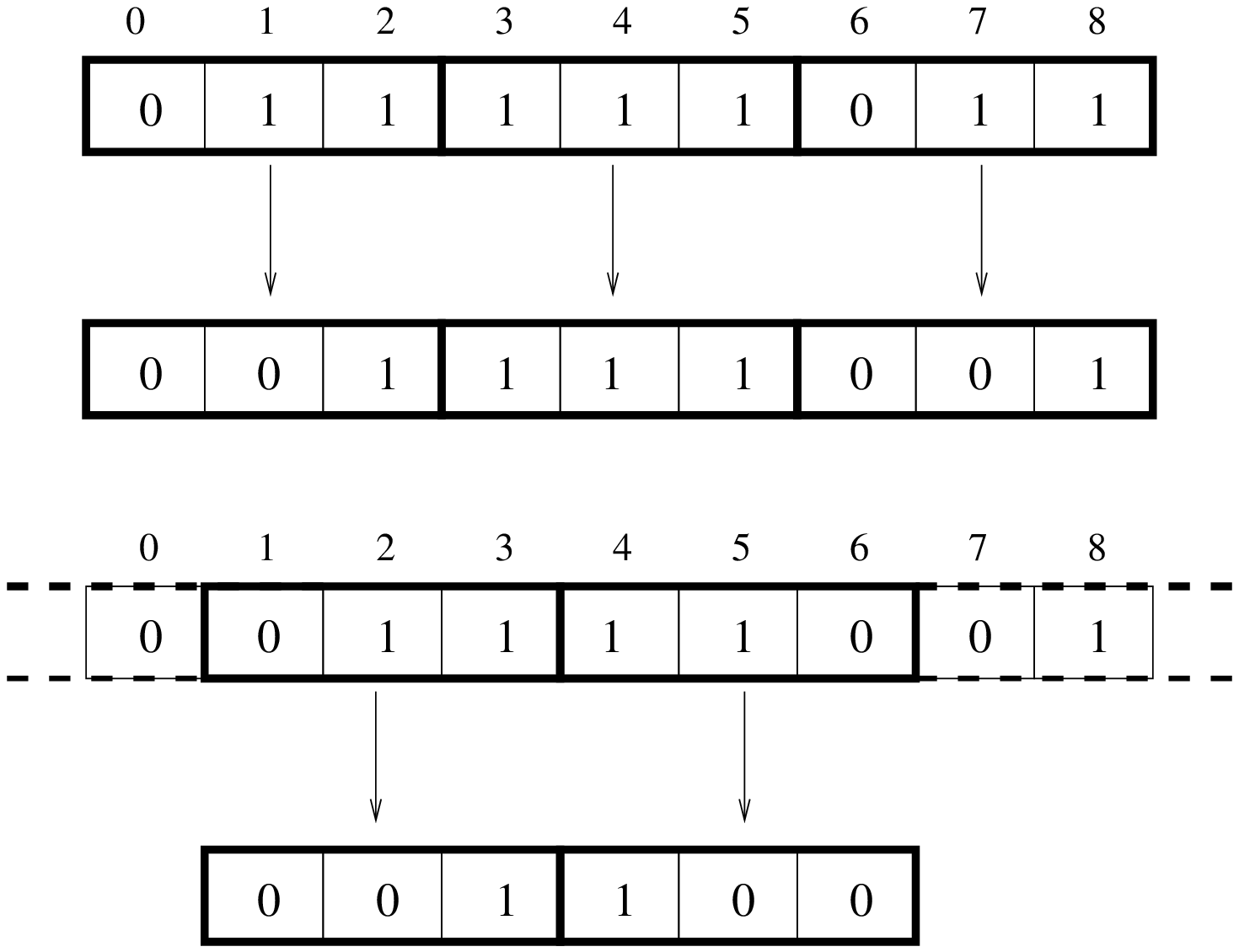, width=4.5cm} }
\caption { Example of a BCA where the transition is applied within blocks consisting of three sites.}
\label{BCA}
\end {figure}

  In Fig.~\ref{BCA} we have an example of using
blocks in a very simple one-dimensional BCA. The only reaction rule is
that the state of a site (0 or 1) becomes 0 if at least one of the 
neighboring sites is 0, otherwise it stays the same.

 To avoid the exchange of information between the blocks and the
problems that might appear at the edges, we generalize this idea of
a block to a partition.
  We define a partition $P$ as a collection $P_i$ of subsets 
of $\Omega$ called {\it chunks}, such that these chunks are disjoint and cover
the entire lattice, $\Omega=\sum_i P_i$.
 For instance, we can write the previous example in terms of partitions.
We have in this case two partitions $P$ and $Q$ consisting of three sets each
where $P_0$=$\{$0,1,2$\}$, $P_1$=$\{$3,4,5$\}$, $P_2$=$\{$6,7,8$\}$ and 
$Q_0$=$\{$0,7,8$\}$, $Q_1$=$\{$1,2,3$\}$, $Q_2$=$\{$4,5,6$\}$.

The new definition of a partition introduces more freedom: we can assign non-adjacent sites to the
chunks such that the problem with the edges between the chunks disappears.
 This means that the conflicts 
should disappear and we therefore add the following restriction.

\begin{quote}
{\it  Each site of the lattice is assigned to a chunk such that between the
sites in the same chunk there are no conflicts for the given model.}
\end{quote}
This means that for all  $s, t$ $\in$  $P_i,$ $s \neq t$ and for all
reaction types $Rt$, $Rt^{\prime}$
\[ Nb_{Rt}(s)\cap Nb_{Rt^{\prime}}(t)=\emptyset \]

 As all the sites in a chunk can be simulated simultaneously, we are
interested to minimize the number of these chunks of a partition (i.e., minimize $|P|$) in order to 
increase parallelism.

\begin{center} {\it Example}
\end{center} 

\begin{figure}
\centering
\subfigure {\epsfig {figure=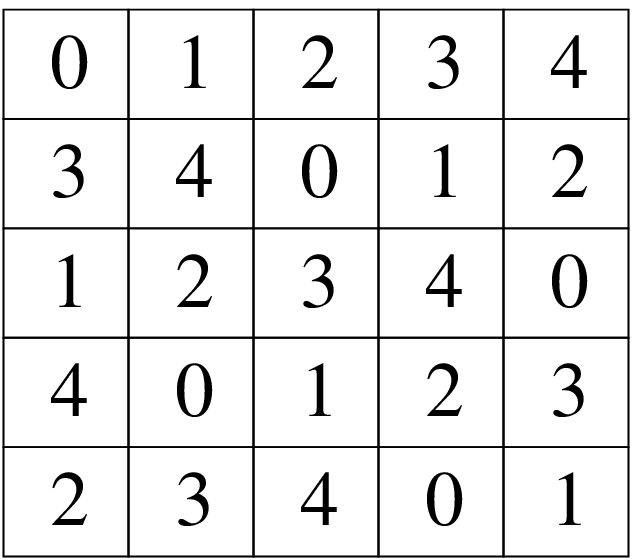, width=2.5cm} }
\caption { The lattice is split into a number of chunks (five in this
case), and the sites are distributed between these chunks. We have
labelled the sites with numbers from 0 to 4, according to the chunk they
belong to. The sites into one chunk can be updated simultaneously, giving
a von Neumann neighborhood.}
\label{partitions_table}
\end {figure}

 We consider again the physical model used in the previous section for $\rm CO$
oxidation on a catalyst surface (see Fig.~\ref{Ziff_model}).
 From Table~\ref{tab:Table1} we can see that the  reaction types do not
include more than two sites.
 When assigning sites to the chunks we observe that a minimum number of 
five chunks can be used, such that the patterns
of the defined reaction types applied at the sites of these chunks do not
overlap.
 In Fig.~\ref{partitions_table} we have a block of 5 x 5 lattice sites,
optimally divided into a number of five chunks. We can use this block
as a pattern to tile the whole lattice.

\begin{center} {\it Algorithm}
\end{center} 

 We give here a NDCA that uses the concept of partitions. 
We call it the Partitioned NDCA algorithm (PNDCA) and it reads as follows
\begin{tabbing}
\hspace{0.2cm} \= for each step\\
               \> ~~ \= choose a partition $P$; \\
               \>    \> for all $P_i \in P$\\
               \>    \> ~~ \= for each site $s {\in} {P_i}$\\
               \>    \>    \> ~~ \= 1. \=  select a reaction type with probability $k_i/K$;\\
               \>    \>    \>    \> 2. \> check if the reaction is enabled at $s$;\\
               \>    \>    \>    \> 3. \> if it is, execute it;\\
               \>    \>    \>    \> 4. \> advance the time;
\end{tabbing}
 The idea is to simulate the enabled reactions according to their rate constants
visiting all the sites of the chunks $P_i$.

\subsection*{Opportunities for improvements}

  The RSM simulations differ from PNDCA simulations for the following
reasons. 
In a PNDCA, in a step, a site can be selected only once for the simulation
because a chunk is selected once per step, and in each chunk a site
is selected exactly once.
After the site has been selected once, the probability to select that
site again in the same step is 0. In RSM there is a probability 1/$N$ per 
iteration to select a site and a non-zero probability to choose the site again in
the same step. Another reason is that while the reactions in a chunk are simulated, the
enabled reactions in other chunks are postponed for execution.

 The deviation can be made smaller through additional
randomization and through re-organization of the steps in the algorithm.
 Depending on how the selection of a chunk is done, we can derive a set
of algorithms. Chunks can be selected in the following ways:
\begin{enumerate}
\item all chunks in a predefined order,
\item all chunks randomly ordered,
\item a set of random chunks such that a chunk has a probability $1/|P|$
to be selected during a step,
\item a weighted selection according to the rates of enabled reactions in each chunk.
\end{enumerate}
Simulating all the chunks per step in order or randomly,
introduces correlations in the occupancy of the sites. More correlations
between occupancy of the sites occur as less chunks are introduced. If $|P|$
is large the algorithm performs better. If $|P|=N$ and a chunk is selected
randomly, PNDCA and RSM match.

We can also vary the amount of work done per chunk through the choice of the number of trials,
$L$. If this number is small only a relatively small amount of time is spent within
the chunk resulting in a small overall deviation. This leads to the following general structure.
\begin{tabbing}
\hspace{0.2cm} \= for each step\\
               \> ~~ \= choose a partition $P$; \\
               \>    \> set {\em trials} to 0;\\
               \>    \> repeat \\
               \>    \> ~~ \= select $P_i \in P ~~$(probability $|P_i|/|P|)$;\\
               \>    \>    \> select $L, 1 \leq L \leq (N-trials)$ \\
               \>    \>    \> set {\em trials} to {\em trials} + L;\\
               \>    \>    \> for $L$ sites $\in {P_i}$\\
               \>    \>    \> ~~ \= 1. \= select a reaction type with probability $k_i/K$;\\
               \>    \>    \>    \> 2. \> check if the reaction is enabled at the site;\\
               \>    \>    \>    \> 3. \> if it is, execute it;\\
               \>    \>    \>    \> 4. \> advance the time; \\
               \>    \> until {\em trials} = N
\end{tabbing}
The number of trials per chunk is smaller than the size of a chunk, $L \le
|P_i|$, and the sites to be visited in a chunk are selected randomly.
We name this algorithm L-PNDCA.
Through special choices of the parameters that we have introduced the L-PNDCA approaches the DMC method
RSM. For example, when $L$ is fixed at 1 or when $|P|$ assumes the extreme values 1 or $N$.

\subsection*{Another approach using partitions}

 The effect of the non-overlap rule is that the patterns in the model limit
the choice of the partition significantly: larger patterns lead to more 
chunks. Since the degree of concurrency is related to the size of
the chunks the non-overlap rule also limits the concurrency.
 By adding an additional ordering constraint we can reduce this effect as
follows.

If we look at the simulation of a single chunk in the partition then this
simulation proceeds by repeatedly selecting a reaction type and executing it.
 We re-order these steps by partitioning the set of reaction types $\cal T$ into 
$\sum_j {\cal T}_j$. The sets ${\cal T}_j$ are selected according to their rates and then the
algorithm is executed for the reaction types in this selected ${\cal T}_j$.
 We can now do a partitioning of the set $\Omega$x${\cal T}$ = $\sum$ ($P_i$,
${\cal T}_j$). 
The non-overlap rule reduces to non-overlap with respect to the reactions 
types within ${\cal T}_j$ and, as a result, the partition can be done with fewer 
chunks. There is a trade-off however: the work per chunk is less, in principle.

\begin{figure}
\centering
\subfigure {\epsfig {figure=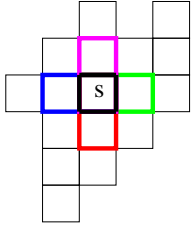, width=2.5cm} }
\caption { Overlap of the reaction patterns on the central site $s$ for the
model of CO oxidation.}
\label{pair_conflict}
\end {figure}       

\begin{figure}
\centering
\subfigure {\epsfig {figure=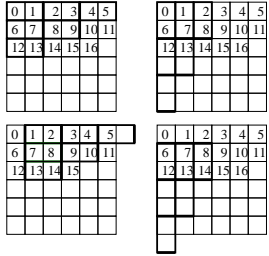, width=3.5cm} }
\caption { The four partitions of the lattice sites according to the  subsets $T_i$
of reaction types, for the model of CO oxidation.}
\label{pairs}
\end {figure}       

\begin{center} {\it Example}
\end{center} 

 The patterns of the  reaction types that can be enabled at a specific site
$s$, all contain two sites positioned as in Fig.~\ref{pair_conflict} and the site $s$
can thus be a part of four possible pairs during a transformation.
 We split the set of reaction types in subsets, such that the patterns of
the reaction types from a subset $T_i$ are included in only one such pair, apart
from translation.
 The set of reaction types is then a collection of two subsets ${\cal T}_j$,
$\cal T$=$\sum_{j=0}^{1} {{\cal T}_j}$ (see Table~\ref{tab:Table2}).
The partition $P$ is a collection of only two chunks
which are constructed by skipping a row or a column each time.
Fig.~\ref{pairs} illustrates the principle for our earlier example.
 The chunks $P_i$ are the following
\\
$P_0$=$\{$0, 2, 4, 7, 9, \ldots$\}$\\
$P_1$=$\{$1, 3, 5, 6, 8, \ldots$\}.$\\

\begin {table}
\begin {center}
\begin{tabular}{|l|p{3.0cm}p{1.5cm}p{1.5cm}|}
\hline  $T_0$ & $Rt_{\rm CO+O}^{(0)}$, $Rt_{\rm CO+O}^{(2)}$ & $Rt_{\rm O_2}^{(0)}$ & $Rt_{\rm CO}$\\
   $T_1$ & $Rt_{\rm CO+O}^{(1)}$,$Rt_{\rm CO+O}^{(3)}$  & $Rt_{\rm O_2}^{(1)}$  & - \\
\hline
\end{tabular}
\end{center}
\caption {\label{tab:Table2} The new division of the reaction types in
subsets ${\cal T}_j$, $j=0,1$ for the model in Fig.~\ref{Ziff_model} applied on
 a site $s$.}
\end{table}

%
%

\begin{center} {\it Algorithm}
\end{center} 

 The algorithm consists of the following steps

\begin{tabbing}
\hspace{0.2cm} \= for each step\\
               \> ~~ \= for $|{\cal T}|$ times\\
               \> \> ~~ \= select ${\cal T}_j$ $\in {\cal T}$ with probability $K_{{\cal T}_j}/K$;\\
               \> \> \>  select a reaction type from ${\cal T}_j$ with probability $k_i/k_{{\cal T}_j}$;\\
               \> \> \>  select $P_i \in P$\\
               \> \> \> for each site $s {\in} {P_i}$\\
               \> \> \> ~~ \=  1. check if the reaction is enabled at $s$;\\
               \> \>  \> \>  2. if it is, execute it;\\
               \> \>  \> \>  3.  advance the time;\\
\end{tabbing}


 This is basically the generalization of the simulation algorithm used by
Kortl$\ddot{\rm u}$ke~\cite{kortluke}.
 We have used here $K_{{\cal T}_j}$ to denote the sum of the reaction rates of the
elements from the set ${\cal T}_j$.

\section* {6. Correctness and performance}

  In order to study the correctness of our methods, we compare the kinetics of
the reactions in DMC and in CA simulations.
  DMC is based on the following fundamental assumption known as Gillespie
hypothesis:

\begin{quote}
  {\it If a reaction with a rate constant $k$ is enabled in a state $S$, then
the probability of this reaction occuring in an infinitesimal time interval
of size ${\delta}t$ is equal to $k {\delta} t$. The probability that more
than one reaction occurs in an interval of length $\delta t$ is negligible.
}
\end{quote}

 The above assumption says that in any state and at any time, the
probability of occurence of an enabled reaction in a vanishingly small
interval is proportional to its rate, and that the probability of two
reactions occuring simultaneously is negligible~\cite{gillespie1,
gillespie2, jansen1}.
 A stochastic model that respects the fundamental assumption is described by
a Master Equation and can be simulated using DMC
methods~\cite{kampen,b_g_s_j, gelten_santen_jansen, jansen, gelten_all, 
lukkien, binder}.

 Based on the Gillespie hypothesis, Segers et al.~\cite{segers} has given two criteria
for the correctness of the simulation algorithms of surface reactions. These
criteria suggest a way to select a site and a reaction type with a correct
probability. According to these criteria, each algorithm is correct if only
enabled reactions are performed and 2 conditions are satisfied: 
\begin{enumerate}
\item  the waiting time for a reaction of type $i$ (the time that elapses before
it occurs) has an exponential probability distribution ($\exp(-k_it)$);
\item the waiting time of the next reaction type $i$ is according to
the ratio between the reaction rate constant ($k_i$) and the sum of the rate
constants of all the enabled reactions.
\end{enumerate}
 In our CA methods with partitions, we have seen that the order of visiting
the sites is important and can introduce correlations in the
simulations. Using partitions, some sites are excluded from simulation for a
certain time, while others are preferred. 
 The same problem arises for executing the enabled reactions. Executing reaction
types from a chunk, disables the enabled reactions in other
chunks and introduces biases in the rates of reactions.
 This comes from the fact that for CA methods, the fundamental assumption
does not hold. In principle, many reactions can be enabled and executed
during a small time interval in a CA. Thus, CA gives results that may deviate
from the DMC results.


  As an example for our new methods, we consider the model used by Kuzovkov et al.~\cite{kuzovkov1} to study reactions on the surface including surface reconstruction.
 The model used is similar to our example model, the oxidation of CO on a face of
Platinum(100). Adsorbates like CO can lift the reconstruction of the
hexagonal structure of the top layer of Pt(100) to a square structure.
 CO adsorbs in both phases of the top layer. $\rm O_2$ adsorbes only in the square phase.
Adsorbed CO may desorb again and O and CO may desorb associatively, forming
$\rm C O_2$. The behavior on the surface is the following: CO adsorbs on Pt(100)
in a hexagonal phase, the surface top layer reconstructs into a square
structure such that $\rm O_2$ can now adsorb on the lattice. As $\rm O_2$
molecules are adsorbed, $\rm CO_2$ is produced and desorb liberating the lattice
from particles. The surface reconstructs again to a hexagonal structure and
the process is repeating: we get oscillatory behavior on the
surface. We use the oscillations in the coverages with particles of the
lattice for comparing our results.

\begin{figure}[h]
\centering
\subfigure {\epsfig {figure=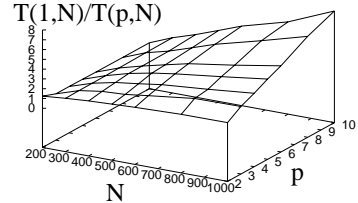, width=5.5cm} }
\caption{ Speedup function on the system size $N$ and on the number of processors
$p$.}
\label{speed}
\end {figure}

 Because of the absence of conflicts between the chunks, the approach with 
partitions leads to a significant increase in speedup even for not so large systems
(100x100, 200x200). In Fig.~\ref{speed} we can see how the speedup of the
PNDCA algorithm depends on the system size $N$ and on the number
of processors $p$. The speedup is defined as the ratio between the
simulation time using 1 processor, for a system size N (T(1,N)) and the
simulation time on $p$ processors, for a system size N (T($p$,N)).

 For the method with partitions we see that a small time spent within a
chunk (small L) affects the parallelism, while a large time spent within
a chunk (large L) affects the correctness.

\begin{figure}[h]
\centering
\subfigure {\epsfig {figure=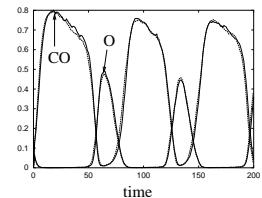, width=3.5cm} }
\caption {RSM and L-PNDCA results for the time dependence of the
coverage with CO and O particles of a system of size $N$=100x100. The
continous line is for RSM, and the dashed line is for L-PNDCA 
with parameters $m=1$, $L=N^2$ and $m=N^2$, $L=1$.}
\label{extreme_part}
\end {figure}

 We denote with $m$ the number of chunks ($m=|P|$).
 In Fig.~\ref{extreme_part} we see that for $m=1$, $L=N^2$ (one chunk
containing all the lattice sites), and for $m=N^2, L=1$ ($N^2$ chunks
with one site per chunk), DMC and $\ \ \ \  $ {L-PNDCA} give the same
results.

\begin{figure}[ht]
\centering
\subfigure {\epsfig {figure=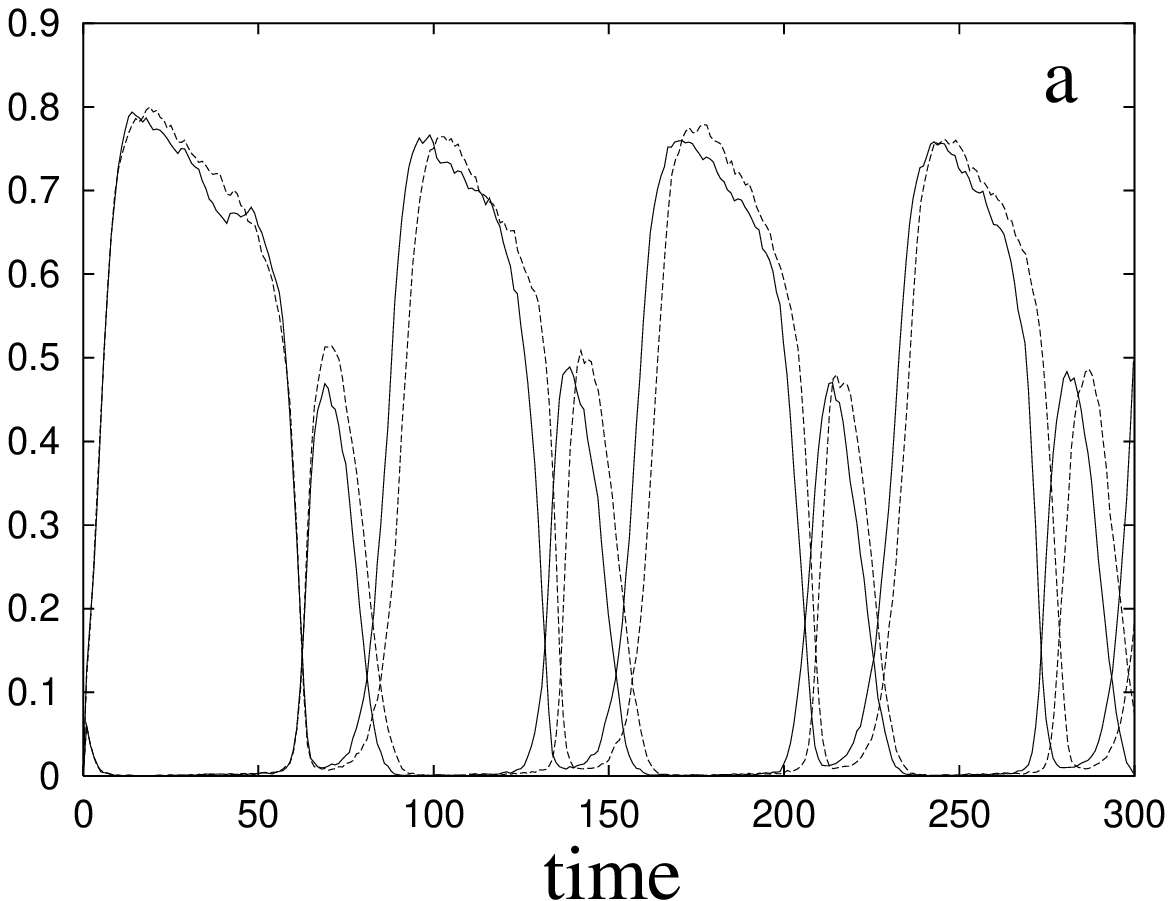, width=3.5cm} }
\subfigure {\epsfig {figure=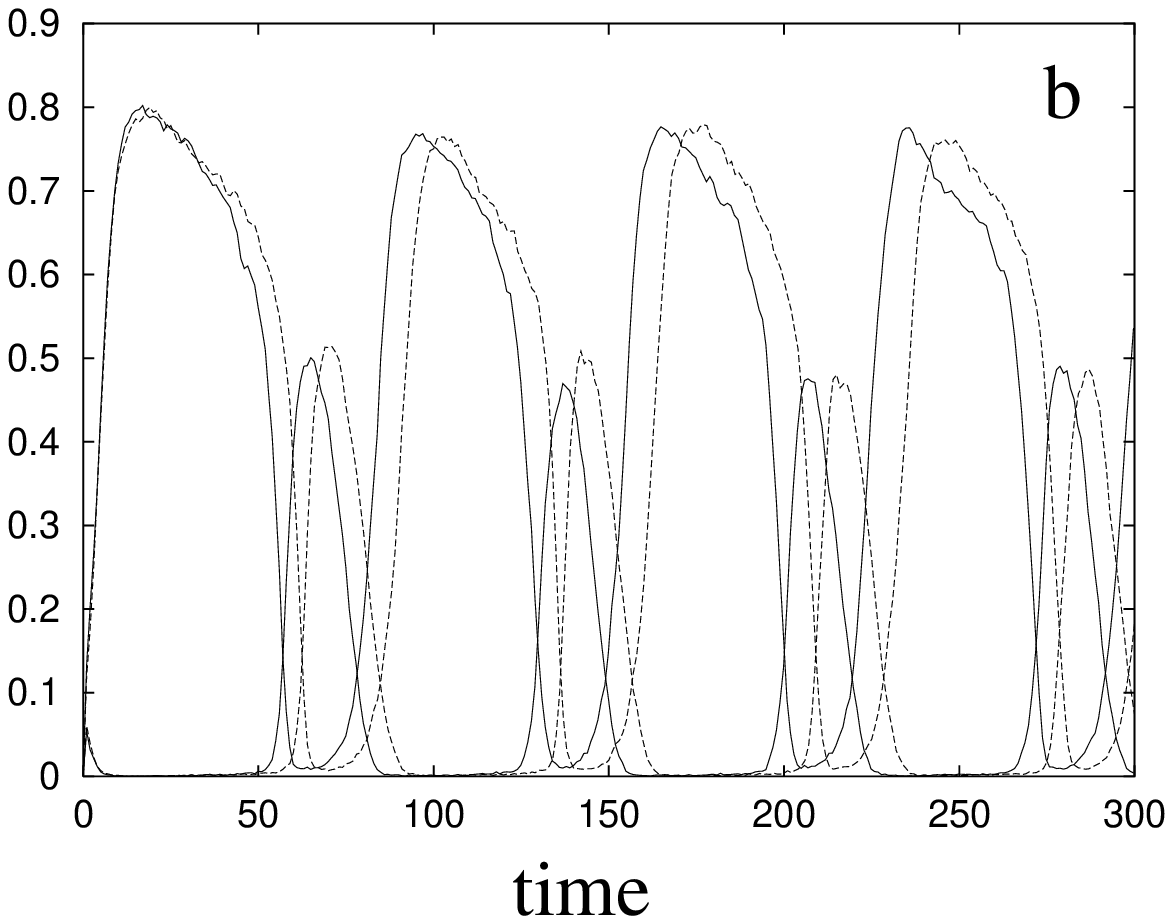, width=3.5cm} }
\caption {RSM and L-PNDCA results for five chunks and $N$=100x100. In a) $L=1$
and in b) $L$=100.}
\label{L_part}
\end {figure}

 For the optimal case of five chunks, we experimented with
different $L$'s. Increasing $L$ introduces biases in the simulations.
In Fig.~\ref{L_part} we illustrate this for case $L=1$ and $L=100$, for
lattice size $N$=100x100.
 We notice that for $L=1$, L-PNDCA gives almost the same results as 
DMC.

\begin{figure}[ht]
\centering
\subfigure {\epsfig {figure=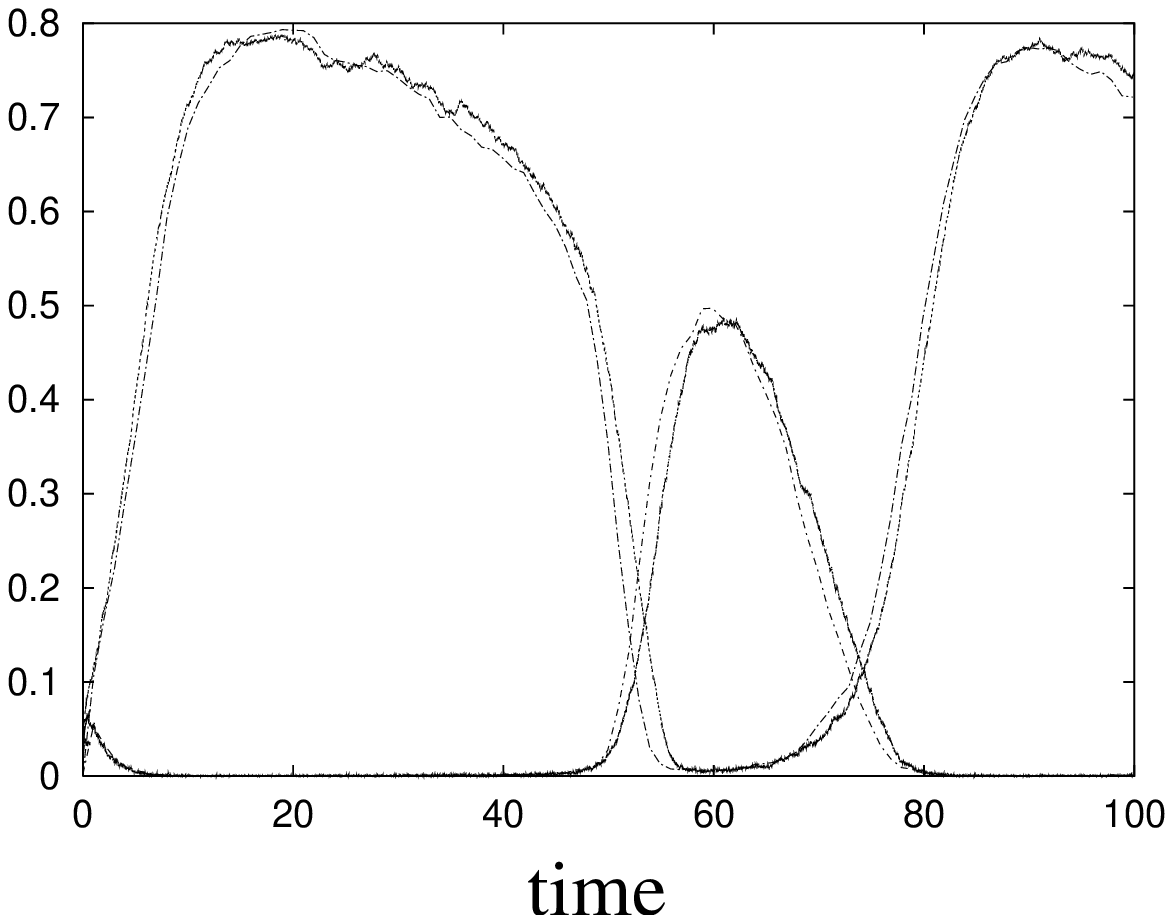, width=3.5cm} }
\caption {RSM and L-PNDCA results for five chunks ($m=5$), $N$=100x100, $L=N^2/m$,
when all the chunks are randomly selected exactly once per step.}
\label{Ln2_5_1}
\end {figure}

 In case a chunk is selected each time with a probability $|P_i|/|P|$,
for larger values of $L$ (e.g. $L$=100), the correlations have as effect
the deviation in time of the oscillations from the DMC results. In this
case, for very large values of $L$, the oscillations disappear.
 But, if we simulate all the chunks only once per step in a random order, 
we get oscillatory behavior even for very large values of $L$
($L=N^2/m$)(see Fig.~\ref{Ln2_5_1}). 
In this case, if we consider very fast diffusion and small probabilities for
chemical reactions in the cells, the deviations are so small that DMC and 
L-PNDCA give similar results. We can have in this case full parallelization 
and very accurate results.

\section*{Conclusions}
 We have presented a collection of approximate algorithms based on the Cellular
Automaton model for parallel simulation of surface reactions. We have
introduced the concept of partitions in a Non-Deterministic Cellular
Automaton, and we have derived a set of parameterized algorithms based on the
partitions concept. The approximation in these algorithms is introduced
through parameters.
 The Cellular Automaton is already an approximate approach that can be taken
as a starting point for simulation of surface reactions, but it gives
results that deviate from the Master Equation.
 These CA algorithms simulate for the limit parameters the Master Equation,
such that the accuracy of the simulations can be compared for different
parameters sets.
 We find that we can get fast simulations using this approach trading accuracy for
performance. We give an example when we can use full parallelization
getting accurate results.
 
\bibliographystyle{./apsrev}
\nocite{*}

\end {document}